\documentclass{article}
\pdfoutput=1

\usepackage{arxiv}

\usepackage[utf8]{inputenc} 
\usepackage[T1]{fontenc}    
\usepackage{hyperref}       
\usepackage{url}            
\usepackage{booktabs}       
\usepackage{amsfonts}       
\usepackage{nicefrac}       
\usepackage{microtype}      
\usepackage{lipsum}
\usepackage{graphicx}
\usepackage{cleveref}
\usepackage{multirow}
\usepackage{adjustbox}
\usepackage{cellspace}
\usepackage{tabularx}
\usepackage{rotating,siunitx}
\usepackage{multirow}
\usepackage{makecell}
\usepackage{longtable}
\usepackage{pdfpages}

\title{PRE-PROCESSING IMAGES USING BRIGHTENING, CLAHE AND RETINEX}

\author{
Thi Phuoc Hanh Nguyen, Zinan Cai, Khanh Nguyen, Sokuntheariddh Keth, Ningyuan Shen, Mira Park \\*
Centre of Digital Technologies Education Research (CODER) \\*
ICT, School of Technology, Environments \& Design \\*
University of Tasmania \\*
\texttt{mira.park@utas.edu.au}
}

\begin{document}
\maketitle
\begin{abstract}
This paper focuses on finding the most optimal pre-processing methods considering three common algorithms for image enhancement: Brightening, CLAHE and Retinex. For the purpose of image training in general, these methods will be combined to find out the most optimal method for image enhancement. We have carried out the research on the different permutation of three methods: Brightening, CLAHE and Retinex. The evaluation is based on Canny Edge detection applied to all processed images. Then the sharpness of objects will be justified by true positive pixels number in comparison between images. After using different number combinations pre-processing functions on images, CLAHE proves to be the most effective in edges improvement, Brightening does not show much effect on the edges enhancement, and the Retinex even reduces the sharpness of images and shows little contribution on images enhancement.
\end{abstract}

\section{Introduction}
Pre-processing attracts great attentions from researchers as well as many practical projects in which the pre-processing  helps to provide high quality of images for analytics. Those methods can enhance the images of biological projects as well. For instance, in the paper ‘Biologically inspired image enhancement based on Retinex’, they illustrate images process method including four steps: illumination estimation, reflection extraction, color-restoration and post-processing. Illumination is based on an available guidance on filter. The reflection step is subject to Retinex algorithm. They propose to use a method named as the modified Contrast–Naturalness–Colorfulness (MCNC) function using the optimal parameters of non-linear stretching to improve the image quality measurement \cite{vincent2009descriptive}. Other papers ‘Observation of deep seafloor by autonomous underwater vehicle’  \cite{ura2013observation} and 'Enhancement of deep-sea floor images obtained by an underwater vehicle and its evaluation by crab recognition' \cite{ahn2017enhancement} explain the importance of using autonomous underwater vehicles to enhance images quality and highlight some relevant methods used for pre-processing images. They raise common issues with undersea images that are the inconsistency between images in different locations and altitudes. A number of methods have been proposed for underwater images to improve the light attenuation, deal with the inconsistency in images quality using homomorphic filtering to remove the effects of non-uniform illumination and some other methods to reduce noises, enhance edges, equal RGB to adjust dominant colours \cite{ahn2017enhancement}. Sueyang \cite{fu2014retinex} considers the three most common issues of undersea images of colour distortion, under-exposure and fuzziness, the paper proposes a Retinex-based method comprising of three steps: colour correction to adjust the distortion in colours, a variational framework for Retinex to improve brightness and image details, the last step is to improve the reflectance and the illumination by a different strategy. There are many research study for finding the best pre-processing methods or the best combination of pre-processing methods.

\section{Methods}
\label{sec:headings}
We have proposed a set of processing steps including the following pre-processing functions:  
\begin{itemize}
    \item Redial Brightening which is based on the suggested parameters 
    \item Contrast Limited Adaptive Histogram Equalization (CLAHE) 
    \item Retinex (Multi Scale Retinex with Colour Restoration and Single-Scale Retinex) 
\end{itemize}
     
These methods are separately applied to images and combined with each other to image processing in different sets. Following is the illustration of the each of Brightening, CLAHE and Retinex.

\subsection{Radial brightening}
The function used in our experiment is based on the recommended algorithm according to Mehrnejad et al \cite{mehrnejad2014towards}. According to them, the value \((V_{new})\) of an image in HSV is represented based on the value of the point in the middle bottom of the picture \(V_{old}\) and the distance of a single point in the image towards the middle bottom point \((D)\) as the following:  
\begin{equation}
V_{new}(x,y) = V_{old}(x,y) + KD(x,y,x_0,y_0)
\end{equation}

\textit{K} in the given equation is chosen as 0.00025 which is chosen through trials \cite{mehrnejad2014towards}. This means that the value of each pixel at \((x,y)\) in the HSV-based image increases subject to the pixel distance of a particular point to the middle point in the bottom image.

The given single Brightening method does not show much difference from the original one. 

\subsection{Single-Scale Retinex (SCR)}
Retinex algorithm \cite{parthasarathy2012automated}, in mathematics, is represented as: 

\begin{equation}
R(x,y) = \log(I(x,y) - \log[F(x,y)*I(x,y)]
\end{equation}

where \(*\) is the convolution operation, \(R(x,y)\) is the associated Retinex output, \(F(x,y)\) is a Gaussian surround function which is represented as:

\begin{equation}
F(x, y) = \epsilon^{-r^2/c^2}
\end{equation}

For the Gaussian function, \(c\) is the dynamic range compression which is sacrificed to improve the rendition. It means that we cannot achieve \(F(x,y)\) and \(c\) of high value at a time. This is a drawback of Single-Scale Retinex and multiple scale Retinex is born to deal with that \cite{jobson1997multiscale}.

\subsection{Multi scale Retinex with Colour restoration (MSCR)}

Retinex algorithm makes images to be ‘greying out’ which means that all three bands of an image are the same after the processing and turns the images into a desaturated image which can be severe in many cases \cite{parthasarathy2012automated}. The proposed restoration algorithm in that paper suggests the function to avoid the side effect of Retinex.

The algorithm for colour restoration is given below: 

\begin{equation}
R(MSRCRi)(x, y) = G(Ci(x, y)R(MSRi)(x, y) + b
\end{equation}

Where \(Ci(x, y) = f[Ci(x, y)]\) is the i\textsuperscript{th} band of the colour restoration function (CRF) and RMSRCRi is the i\textsuperscript{th} spectral band of the Multiscale Retinex with colour restoration.

\begin{equation}
C_i(x, y) = \beta\log[\alpha I'_i(x,y)] = \beta \log[\alpha I_i(x,y)] - \beta [\Sigma^S_{i=1}I_i(x,y)]
\end{equation}

where \(\beta\) is a gained constant, \(\alpha\) controls the strength of the non-linearity, \(G\) and \(b\) are final gain and offset values. The values specified for these constants suggested in \cite{jobson1997multiscale} are \(\beta\) = 46, \(\alpha\) = 125, \(b\) = -30, \(G\) = 192. On the implementation the algorithm, we utilise the suggested parameters from \cite{jobson1997multiscale}.

The algorithm helps to enhance images by using a wide range of nonlinear illumination conditions. The algorithm uses different parameters subject to the consideration on different kinds of images. Based on a research by Parthasarathy \cite{parthasarathy2012automated}, which illustrates the instructions to find out proper parameters automatically. The paper also admits the failure of the Multi Scale Retinex with Colour Restoration as it still results in greyed out images. 

\subsection{Contrast Limited Adaptive Histogram Equalization (CLAHE)}

CLAHE has been widely used for image enhancement on a histogram basis \cite{reza2004realization}. To be more specific, the algorithm is originated from the assumption that the consistency in all areas within an image is maintained. All those areas on the image will be enhanced by one unique grayscale mapping. However, the distribution of grayscales varies which means there needs to have a method to equalize the image grayscale distribution. Adaptive histogram equalization method supports the finding of the mapping for each pixel subject to its neighbour population of grayscale. As each pixel gets the calculation of contrast enhancement mapping, the number of times that method is repeated is the number of pixels in an image which means that the method requires an extensive computational basis \cite{reza2004realization}. 

Technically, CLAHE does this by setting a threshold. If some grey levels in the image exceed the threshold, the excess is evenly distributed to all grey levels. After this processing, the image will not be over-enhanced, and the problem of noise amplification can be reduced \cite{pizer1987adaptive}.

For the blocky area problem of Adaptive Hisogram Equalization, \cite{pizer1987adaptive} proposes an interpolation algorithm. By increasing the number of mapping function values of each pixel, the method balances the differences in grey values of adjacent local image processing blocks.

OpenCV provides the CLAHE function, which has two parameters: \(clipLimit\) and \(tileGridSize\). \(clipLimit\) represents the threshold clip size mentioned above and \(tileGridSize\) represents the size of the image processing window.

After applying three functions of pre-processing in different ways, Canny Edge detection is utilised to evaluate the edge detection extent of different sets of preprocessing methods. Canny Edge detection algorithm represents a two-direction spatial measurement in images. Canny Edge as an edge detector uses two 3x3 convolution masks with x-direction and y-direction illustrate estimating gradient. Canny Edge detection is said to be highly sensitive with noises and highly recommended for data communication and data transfer \cite{reza2004realization}, this is also the reason why we opt for Canny Edge detection for image evaluation to effectively highlight noises as edges as it was found.

\subsection{Implementation}

We use python for the implementation of our experiments. For the Retinex method, we refer to the code of \url{https://github.com/dongb5/Retinex} and choose the same parameters. For the CLAHE method, we use the CLAHE function that comes with OpenCV. In the selection of parameters, after comparing different image results, we set the clip limit to 2 and tile grid size to 50. In the Brightening method, the middle point in the bottom of image is taken as a standard for image enhancement process \cite{mehrnejad2014towards}.

Because the order of methods will produce different results, we test all possible combination of those algorithms. In the end, the following 15 different image processing algorithm combinations can be obtained: Retinex, CLAHE, Brightening, Retinex – CLAHE, CLAHE – Retinex, CLAHE – Brightening, Brightening – CLAHE,  Retinex – Brightening, Brightening – Retinex, Brightening – CLAHE – Retinex, Brightening – Retinex – CLAHE, CLAHE – Retinex – Brightening, CLAHE – Brightening – Retinex, Retinex – CLAHE – Brightening and Retinex - Brightening - CLAHE.

The dataset called ‘P81’ is provided to this study by the Institute of Marine and Antarctic Studies (IMAS). The P81 is originated from the paper by Piepenburg \cite{piepenburg2016seabed} where it were taken on both sides of the northern Antarctic Peninsular between January to March 2013 at a depth between 35 to 780m. Among the images in the dataset, there are a number of species such as Amphipods, Bony fish, Brittlestars, Cephalopods, Chitons, Echiura, Seastars and many other species. However, this study will be focusing only on starfish. Therefore, five starfish contained images are chosen including: image 0, image 1, image 5, image 7 and image 9. At the same time, to facilitate subsequent comparisons, we crop these pictures according to the target positions in red squares shown in Figure 1, and finally get 13 images containing starfish. We number the cropped images under the form of i<number of image>\_crop\_<number of crop>. For instance, The Image 0 contains the starfish crop number 1 then the cropped images will be named as i0\_crop\_1.

\begin{figure}
\begin{center}
\begin{tabular}{|c|c|}
\hline
\begin{adjustbox}{width=0.66\textwidth, valign=t}
\begin{tabular}{c|c}
  
  Image 0 & Image 1 \\
  \hline
  \includegraphics[width=0.3\linewidth]{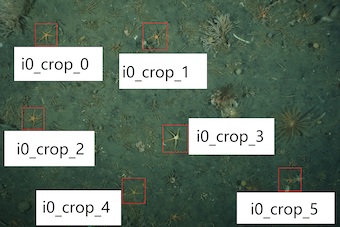} &
  \includegraphics[width=0.3\linewidth]{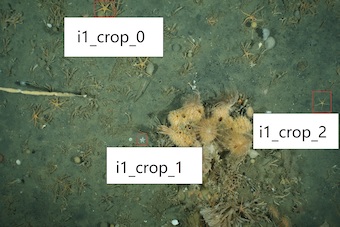} \\
  \hline
  Image 5 & Image 7 \\ 
  \hline
  \includegraphics[width=0.3\linewidth]{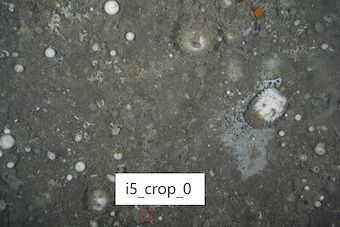} &
  \includegraphics[width=0.3\linewidth]{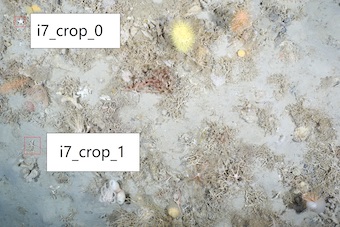} \\

\end{tabular}
 \end{adjustbox}
&
\begin{adjustbox}{width=0.33\textwidth, valign=t}
\begin{tabular}{c}
    \hline
     Image 9 \\
    \hline
  \includegraphics[width=0.3\linewidth]{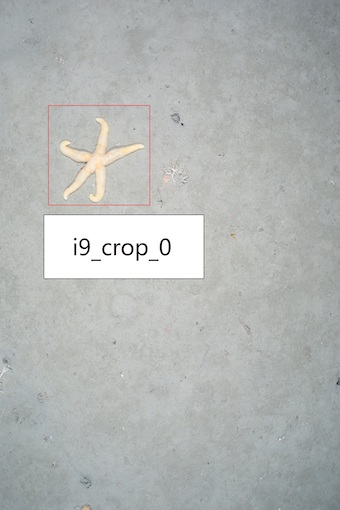} \\
\end{tabular}
\end{adjustbox}
\\
\hline
\end{tabular}
\end{center}
\caption{Starfish crops position in image 0, image 1, image 5, image 7, image 9 as red squares} 
\label{fig:starfish}

\end{figure}

\section{Evaluation}
After we perform the above 15 sets of pre-processing on the 13 samples, we use Canny Edge detection function that comes from OpenCV to process the images, because the image processed using Canny Edge detection, it can quantify the clarity of starfish edges. Before comparing, we need to exclude the interference of irrelevant pixels around the starfish. Therefore, we use \(boundingRect\) (OpenCV image editing component) to manually create a mask based on the shape of the starfish. By blackening the pixels around the edge of the starfish, we can ensure that the true positive pixels counted in the image are almost on the contour line of the starfish. Finally, we can calculate the number of true positive pixels of object images. 

\begin{figure}[h]
\centering
\includegraphics{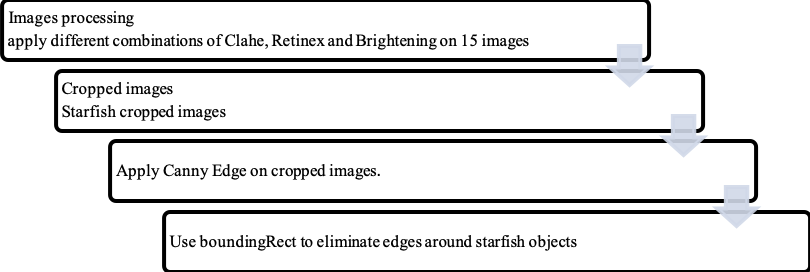}
\caption{Overview of the system.}
\label{fig:overview}
\end{figure}

The images processing procedure is summarised into Figure \ref{fig:overview}. 

The program has been uploaded to GitHub for further research. Please refer to this \url{https://github.com/TsaiZinan/CoDER_Project_Code}

\section{Results}
Appendix A illustrates the cropped processed images after applying Canny Edge detection on. Following is the Table 1 which is the true positive pixels calculation of those crops after above-mentioned four steps in the Figure 1. Of all 13 cropped images, there are 06 pre-processed images with CLAHE that have the biggest number of true positive pixels, 04 images get the biggest number of true positive pixels when applying Brightening and CLAHE and 03 images get the biggest number of true positive pixels after CLAHE processed and Brightening pre-processing. This proves pre-processing sets that include CLAHE, in general, are the most effective in adjusting the sharpness of images. 

When compare the original images with brightening images, the number of true positive pixels increases slightly which is also the case for Retinex-Brightening and Retinex only. The Brightening pre-processing, in conclusion, shows little effect on image sharpness enhancement. 

The Retinex does not help much in increasing the sharpness. Each methods combination including Retinex will significantly reduce the number of true positive pixels. In combination with Retinex, the Brightening even reduces the sharpness of images showing by the true positive pixels of Retinex images ranging from 200 to around 400 all fall down to almost 0. The single Retinex images are all of lower true positive pixels compared to the images processed by other methods.

Order of pre-processing methods within a combination make differences in defining the number of true positive pixels, for example, when apply Brightening then Retinex on an image, the true positive pixels number counted is different from applying Brightening after Retinex or the different orders within Brightening, Retinex, CLAHE result in difference in true positive pixels numbers.

\includepdf[width=\paperwidth]{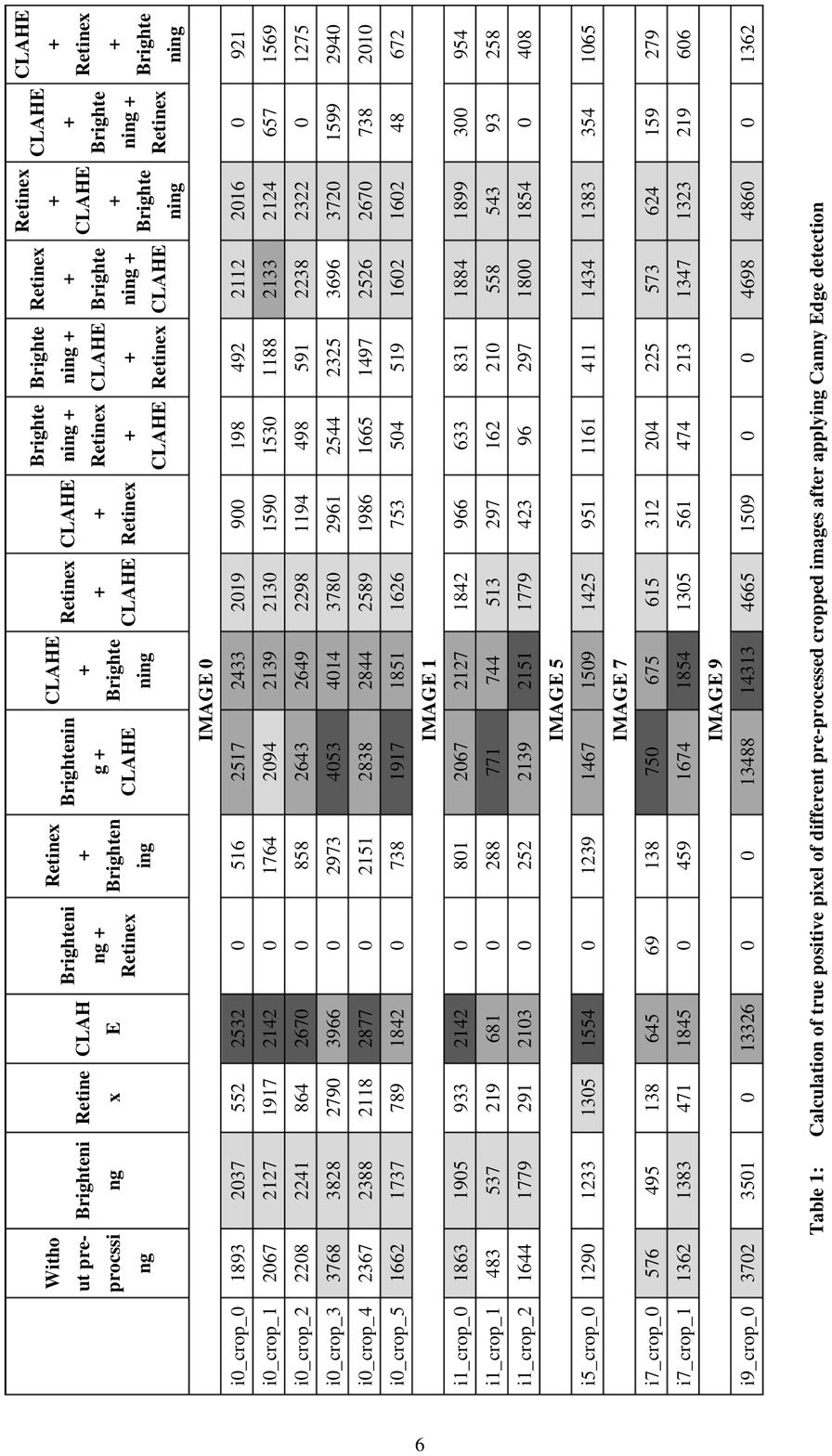}

\section{Conclusion}
\begin{figure}[ht]
\centering
\includegraphics{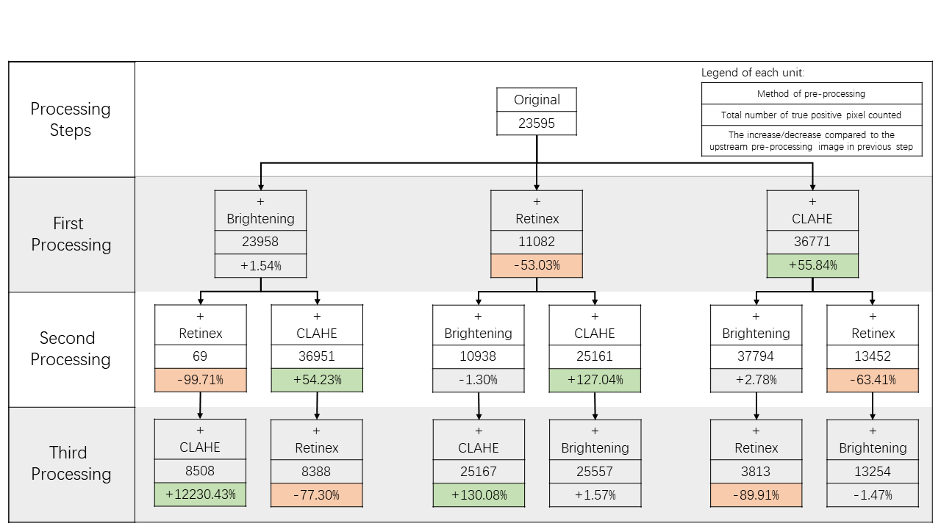}
\caption{Sum of true positive pixels calculation of all pre-processed cropped images by each step.}
\label{fig:true_positive}
\end{figure}

Table 1 can be converted to the flow chart as Figure \ref{fig:true_positive} based on the different sequence of methods using. The first row of Figure \ref{fig:true_positive} shows the total number of true positive pixels in the original image. The second row, which is "First Processing", shows the result after using three different single image pre-processing methods. The third row, "Second Processing", indicates the calculation result of true positive pixels after applying the second processing method with the previous one. And the last row shows the result after the image in the previous row implement the third method. 

In each small block of Figure \ref{fig:true_positive}: the first row represents the method of pre-processing in this processing step; the second row is the sum of true positive pixels in each sample images; compared to the result of the previous processing step, the percentage change in the number of true positive pixels are shown in the third row. Each colour represents a method we use. 

According to the result, it is clear to see that different sequence of methods using does cause different results. In addition, we can find that after implementing Retinex method (orange colour), all the number of true positive pixels decrease from -53.03\% to -99.71\%. All the number after using CLAHE (green colour) increase from 54.23\% to 12230.43\%. With the brightening method (grey colour), the number changes slightly from +2.78\% to -1.47\%.

In conclusion, the CLAHE function can significantly increase the effectiveness of canny edge detection. In contrast, Retinex method dramatically reduces the sharpness of images. And the brightening process does not change the result substantially.

There are some improvements towards image the pre-processing process in further work. Firstly, a larger number of images should be used to ensure credibility. Secondly, the project uses \(boundingRect\), an OpenCV image editing component, to manually create a starfish mask. For large dataset, this method is not applicable as it takes time for manual objects' masks creating. Therefore, we should explore whether there is a more automated way to create a mask. Finally, for each 13-megapixel image used in the project, the Brightening method takes around 40 seconds, and the Retinex algorithm takes roughly 3 minutes with a AMD Ryzen 5 2600 6-cores 12 threads CPU. Therefore, if a subsequent use of a dataset which contains a large number of pictures, a high-performance processor is necessary.

\bibliographystyle{unsrt}  
\bibliography{source}  

\includepdf[width=\paperwidth]{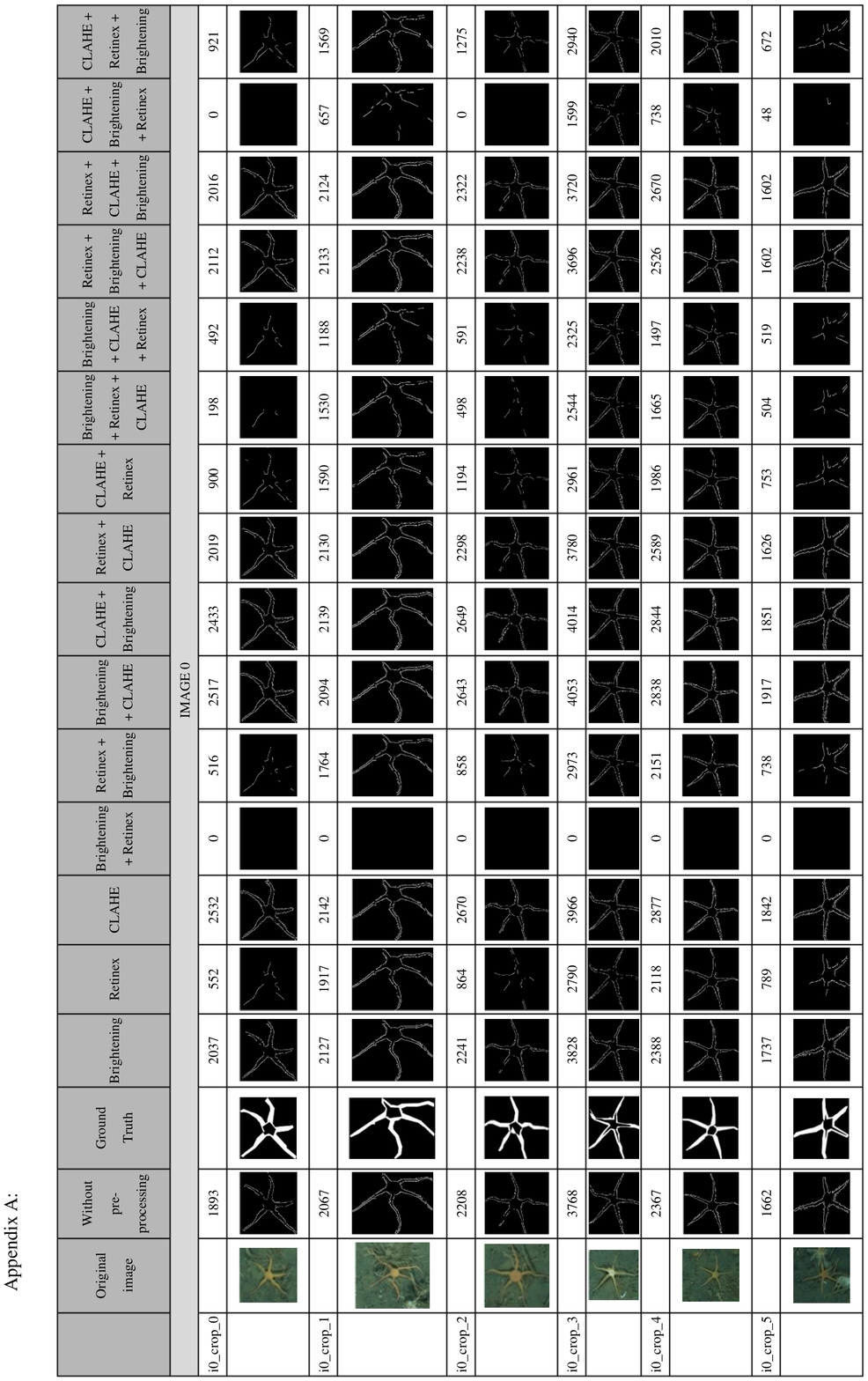}
\includepdf[pages=2,width=\paperwidth]{Appendix.pdf}

\end{document}